# Intrinsic translational symmetry-breaking charge stripes in underdoped iron pnictides

Qiang-Jun Cheng[1], Cong-Cong Lou[1], Yong-Wei Wang[1], Ze-Xian Deng[1], Xu-Cun Ma[1,2,†], Qi-Kun Xue[1,2,3,4,5,†], Can-Li Song[1,2,†]

[1]*Department of Physics and State Key Laboratory of Low-Dimensional Quantum Physics, Tsinghua University, Beijing 100084, China*

[2]*Frontier Science Center for Quantum Information, Beijing 100084, China*

[3]*Shenzhen Institute for Quantum Science and Engineering and Department of Physics, Southern University of Science and Technology, Shenzhen 518055, China*

[4]*Beijing Academy of Quantum Information Sciences, Beijing 100193, China*

[5]*Hefei National Laboratory, Hefei 230088, China*

Despite being well established in cuprates, an intrinsic translational symmetry-breaking charge order has not been clearly identified in iron-based superconductors. Using spectroscopic-imaging scanning tunneling microscopy on epitaxial $Ca(Fe_{1-x}Co_x)_2As_2$ ($x = 0 \sim 0.055$) thin films, we observe smectic, near-commensurate charge-stripe order in the underdoped regime that intervenes between the nematic parent phase and optimally doped superconductivity. Distinct from the bidirectional checkerboard-like order in cuprates, these charge stripes are unidirectional along the antiferromagnetic Fe-Fe bond direction and are accompanied by a van Hove singularity near the Fermi level, inherited from the Fermi surface reconstruction driven by intertwined antiferromagnetic and nematic correlations. Both local and global suppression of the charge-stripe instability enhance superconductivity, tunable via epitaxial strain and Co doping. These results establish charge-stripe order as an intermediate electronic phase in iron pnictides and reveal a coherent pathway from nematicity to superconductivity. Our findings highlight charge ordering as a unifying element across different families of high-temperature superconductors.

[†]To whom correspondence should be addressed. Email: clsong07@mail.tsinghua.edu.cn, xucunma@mail.tsinghua.edu.cn, qkxue@mail.tsinghua.edu.cn

A hallmark of high-temperature ($T_c$) superconductivity is its emergence from a complex landscape of symmetry-breaking states[1-4], including antiferromagnetism[5,6], electronic nematicity[7-12], and spin- and charge-density waves (SDW and CDW)[13-19]. Elucidating the microscopic pathway through which superconductivity develops from these correlated phases is central to understanding the mechanism of high-$T_c$ pairing[3,4], yet it remains one of the outstanding challenges in condensed matter physics. Among these symmetry-breaking phases, charge ordering has been well established throughout the underdoped (UD) regime of many cuprates, which often manifests as bidirectional, checkerboard-like modulations in real space[16-19]. Its prominent role in shaping the electronic phase diagram of cuprates has stimulated intense interest[20-22], with CDW correlations discussed as a competitor to superconductivity[23,24], a precursor to Cooper pairing[25-27], or a manifestation of underlying pair-density-wave states in cuprates[28-30]. This close connection has, in turn, motivated extensive efforts to search for analogous charge ordering in other families of high-$T_c$ superconductors – including the recently discovered nickelates[31,32] – but conclusive experimental evidence has remained elusive.

Iron-based superconductors (IBSs), another key family of high-$T_c$ materials, exhibit dome-shaped phase diagrams reminiscent of the cuprates: superconductivity emerges upon suppression of the antiferromagnetic parent state via chemical doping or pressure[33-35]. The antiferromagnetic phase terminates near a quantum critical point[36], where its associated Fermi surface (FS) reconstruction – present throughout the underdoped regime – disappears and superconductivity reaches its maximum[37,38]. Meanwhile, IBSs ubiquitously host an electronic nematic phase with spontaneous rotational-symmetry breaking below a transition temperature $T_N$[7-10,39], a phenomenon increasingly identified in cuprates as well[11,12,40]. Despite these similarities, and in sharp contrast to structurally related compounds such as $BaNi_2As_2$ that host multiple charge-ordered states[41,42], it remains unresolved whether an intrinsic charge-ordered phase exists in IBSs. This raises the fundamental question of whether charge ordering is generic to correlated high-$T_c$ superconductors and what role, if so, it plays in superconductivity of IBSs.

Spectroscopic-imaging scanning tunneling microscopy (SI-STM), with its ultrahigh spatial and energy resolution, is a powerful probe of electronic symmetry breaking and has visualized unidirectional electronic nanostructures induced by individual impurities in the nematic phase of IBSs[43,44]. Here, we use SI-STM to systematically study the electronic evolution of high-quality epitaxial $Ca(Fe_{1-x}Co_x)_2As_2$ (CFCA) films (Fig. 1a) grown by molecular beam epitaxy (see **Methods**). In the underdoped regime, we directly visualize a smectic, near-commensurate charge-stripe order on the FeAs plane that intervenes between the nematic parent state and superconducting phase (Fig. 1b). This intrinsic translational symmetry-breaking charge stripe is distinct from previously reported charge modulations in stoichiometric FeSe[45], LiFeAs[46], and heavily hole-doped $Ba_{1-x}K_xFe_2As_2$[47] – phenomena that are either strain-induced or confined to surfaces. By harnessing epitaxial strain as an independent control parameter alongside chemical doping (Fig. 1b), we further reveal that the charge-stripe order is correlation-driven, arising from the intertwined antiferromagnetic and nematic correlations inherent to IBSs.

Large-scale STM topography, combined with X-ray diffraction, establishes the high crystalline quality of the epitaxial CFCA films (Supplementary Fig. 1a,b), revealing atomically flat terraces interrupted only by sparse screw dislocations. A uniform step height of 5.7 – 5.8 Å (Supplementary Fig. 1a, inset), corresponding to half the CFCA unit cell (UC), indicates a single, well-defined surface termination. With increasing Co concentration $x$, the out-of-plane lattice constant decreases systematically (Supplementary Fig. 1c), matching with the smaller ionic radius of Co (~ 54.5 pm) relative to Fe (~ 78 pm) and confirming effective chemical substitution. Atomically resolved STM images consistently show an ordered $\sqrt{2} \times \sqrt{2}$ reconstruction (Fig. 1c), characteristic of a polar FeAs termination[48]. This assignment is independently confirmed through controlled Co substitution in the FeAs plane. In addition to intrinsic As vacancies (circled in white) and sparse Ca-for-As substitutions (circled in magenta) in undoped $CaFe_2As_2$ (CFA) (Fig. 1c), Co substitution introduces bright protrusions (circled in black) whose areal density quantitatively matches the nominal Co concentration (for example, ~ 1.7% for $x = 0.015$; Fig. 1d). This one-to-one correspondence across various CFCA samples with variable doping levels (Supplementary Fig. 2) demonstrates that each bright protrusion originates from an individual Co dopant, which locally perturbs the electronic states of adjacent upper As atoms ($As_{up}$) (Fig. 1d inset, black circles). Notably, this termination differs from those reported in $Ba(Fe_{1-x}Co_x)_2As_2$ films[49,50], likely reflecting differences in epitaxial strain (tensile in CFCA versus compressive in Ba-based films) and surface stability of Ca and Ba adatoms. Importantly, these observations provide direct STM access to the FeAs plane – the primary electronic layer hosting superconductivity and electronic orders in 122-type iron pnictides, a termination that is otherwise hard to obtain in cleaved bulk crystals due to their polar and poorly controlled surfaces[51].

Figure 1e displays representative tunneling conductance ($dI/dV$) spectra showing the low-energy density of states (DOS) of 10-UC CFCA films for various Co doping levels ($x$). The parent compound ($x = 0$) exhibits a strongly particle-hole-asymmetric spectrum, dominated by a sharp conductance peak around -5 meV and a broader hump near -15 meV. The sharp DOS peak indicates the presence of a van Hove singularity (VHS), a hallmark of strong electron correlations as previously observed in $KFe_2As_2$[52] and $CaKFe_4As_4$[53]. With slight electron doping (UD, $x = 0.022$), the VHS weakens and shifts to higher binding energy (~ -15 meV) while the overall spectral line shape remains largely intact. Further Co substitution into the optimally doped (OPT, $x = 0.042$) and overdoped (OD, $x = 0.055$) regimes completely suppresses the VHS, giving way to particle-hole-symmetric superconducting gaps (marked by black triangles). The superconducting origin of these gaps is compellingly corroborated by temperature-dependent $dI/dV$ measurements on the optimally doped sample (Fig. 1f), revealing a gap closure temperature (~ 17 K) close to the $T_c$ ~ 14.1 K obtained from macroscopic transport measurements (Supplementary Fig. 3).

We next employ SI-STM to probe the electronic structure on large, atomically flat FeAs surfaces across different Co doping levels (Fig. 2a-c). In the parent compound, conductance maps $g(r, E = eV) \equiv dI/dV(r, V)$ reveal nanoscale nematic domains separated by domain walls aligned along the As-As direction (dashed black lines in Fig. 2a,b and Supplementary Fig. 4a), analogous to those reported in $BaFe_2As_2$ films[49] and consistent with ubiquitous electronic nematicity in IBSs. This nematic state, accompanied by an orthorhombic distortion ($a > b$), is further manifested by unidirectional electronic textures surrounding Ca-for-As substitution defects and As vacancies on the $As_{up}$ layer (Fig. 2a). These dumbbell-shaped features extend over ~$8a_{Fe}$ ($a_{Fe}$ is the nearest-neighbor Fe-Fe spacing) along the orthorhombic $a$ direction, rotate by 90° across the domain walls, and are prominent only near the VHS energy (Supplementary Fig. 4b). Unlike earlier studies that inferred the nematicity through anticorrelation analysis and attributed it to Co dopant-induced electronic nanostructures[54], our measurements directly image the same unidirectional electronic textures linked to intrinsic defects on the FeAs plane.

In the underdoped regime, the electronic structure undergoes a dramatic transformation and reorganizes into uniaxial stripe-like patterns whose orientation rotates by 90° across the nematic domain walls (Fig. 2b). The stripes run along the orthorhombic $b$ axis with a near-commensurate periodicity of ~ $4a_{Fe}$. In some locally strained regimes (Fig. 2b and Supplementary Fig. 5, brown box), the stripe order is significantly suppressed and instead nanoscale superconducting (SC) puddles emerge (Supplementary Fig. 6). This indicates that local strain favors superconductivity at the expense of the stripe order. With further electron doping, both nematic domains and stripe order disappear in the OPT and OD regimes (Fig. 2c and Supplementary Fig. 7), giving way to a phase-coherent superconducting phase. This phase is evidenced by the appearance of Abrikosov vortices under applied magnetic fields (Supplementary Fig. 8a). Pronounced near-zero-energy bound states are observed inside the magnetic vortex cores (Supplementary Fig. 8b), placing the system in the clean limit and confirming the exceptional crystalline quality of the CFCA films. These observations establish a near-commensurate stripe phase that intervenes between the parent nematic state and superconductivity in iron pnictides. Evidently, the stripe order competes with superconductivity locally and globally.

By examining the energy dependence of the stripe-like textures, a pronounced contrast reversal is readily observed in the local DOS across the VHS peak near -15 meV (Fig. 2d), providing direct evidence that the stripe modulations originate from charge ordering. This is corroborated by the stripe-modulated spectra with asymmetric DOS peaks across the VHS energy (Supplementary Fig. 9). To visualize this reversal more clearly, we Fourier-filter a conductance map from the zoom-in stripe region in Fig. 2b (marked by the purple box) by selectively retaining the stripe components near $Q_a \approx 1/4$ ($\pm 2\pi/a_{Fe}$, 0) (see **Methods**)[11]. The corresponding inverse Fourier transform reveals clear stripe modulations whose intensity shifts out of phase across the VHS, further highlighting their short-range, smectic character (Fig. 2e). Consistent with this real-space behavior, the Fourier-transformed maps $g(q, E)$ exhibit stripe-associated peaks only in the underdoped regime. While the peak intensity varies with energy, the wave vectors are essentially non-dispersive and fixed at 1/4 ($\pm 2\pi/a_{Fe}$, 0) (Fig. 2f). This non-dispersive character, along with the contrast reversal and a maximal negative correlation coefficient of -0.24 between $g(r, E)$ acquired above and below the VHS, constitutes definitive evidence for a charge-stripe order in underdoped CFCA. The near-commensurate nature of this charge order is highlighted by overlaying a ball model of the $Fe_2As_2$ layer onto the real-space conductance maps (Fig. 2e).

To elucidate the electronic origin of the emergent charge order in underdoped CFCA, we track the doping evolution of the band structure using quasiparticle interference (QPI) measurements (Fig. 3a-e). Whereas the OPT and OD samples show simple, closed, square-shaped scattering patterns (Fig. 3e and Supplementary Fig. 7b,c), the parent and UD samples display a much richer QPI structure with two distinct scattering channels. One consists of highly anisotropic, spot-like features along the Fe-Fe directions ($Q_a$ and $Q_b$), which are most pronounced at low energies (Fig. 3a,c). The other corresponds to arc-shaped scattering features predominantly distributed along the Γ-M directions ($Q_M$) of the unfolded Brillouin zone (BZ) (Fig. 3b). Notably, the arc-shaped QPI patterns nearly evolve into square-shaped ones in the underdoped regime (Fig. 3d), coexisting with the spot-like feature at $Q_a$. This is more clearly revealed in the high-resolution $g(q, E)$ map derived from the larger $g(r, E)$ map containing multiple domains (Supplementary Fig. 10). For clarity, these dominant QPI features are overlaid onto the corresponding $g(\boldsymbol{q}, E)$ maps in Fig. 3a-e. Analysis of the extracted dispersions for four representative samples reveals a clear dichotomy between the two scattering channels (Fig. 3f,g). The $Q_a$ and $Q_b$ vectors remain essentially non-dispersive over a wide energy range from -70 to -15 meV. Notably, $Q_b$ is strongly suppressed in the underdoped regime, leaving $Q_a$ as the unidirectional scattering vector at $q \approx$ 1/4 ($\pm 2\pi/a_{Fe}$, 0) (Fig. 3c,d). By contrast, the arc- and square-shaped scattering features disperse with energy in all momentum directions, and they originate from the same electron-like bands across the entire phase diagram. This assignment is supported by parabolic fits of the $E$-$q$ dispersions (dashed lines), producing similar band-bottom energies that shift monotonically upward with increasing Co concentration along both the X-Γ-Y (Fig. 3f) and Γ-M (Fig. 3g) directions. This sharp contrast of distinct QPI patterns between the nematic parent state and superconducting state is illustrated in more detail in Supplementary Fig. 11. The dichotomy between the non-dispersive, anisotropic scattering and the dispersive, isotropic scattering strongly implies a Fermi surface reconstruction that crucially depends on the chemical doping.

Previous band calculations and angle-resolved photoemission spectroscopy measurements established that, above the antiferromagnetic transition $T_{SDW} \sim T_N$, the FS of CFCA consists of hole pockets (in red) at Γ and elliptical electron pockets (in blue) at the X/Y points of the unfolded BZ (Fig. 3h)[34-36]. The electron pockets at X(Y) derive from Fe $d_{yz}$ and $d_{xy}$ ($d_{xz}$ and $d_{xy}$) orbitals, whereas the hole pockets primarily originate from the $d_{xz}$ and $d_{yz}$ orbitals. In the high-temperature phase, the $d_{xz}$ and $d_{yz}$ bands remain degenerate, preserving fourfold ($C_4$) rotational symmetry. Below $T_N$, the antiferromagnetic order predominantly folds the electron pockets at X(Y) back to Γ and hybridizes them with the central hole pockets, generating tiny electron pockets (Fig. 3i)[37,38,55,56]. Meanwhile, the nematicity-induced band splitting accompanying these tiny electron pockets contributes large DOS along the Γ-X/Y directions, thus generating the VHS peak near the Fermi energy (Fig.1e)[52,53,57,58]. The reconstructed tiny electron pockets contain nearly parallel segments, giving rise to FS nesting that naturally accounts for the non-dispersive, spot-like $Q_a$ and $Q_b$ features observed in QPI[55]. The anisotropy between $Q_a$ and $Q_b$ arises from nematicity-induced orbital splitting of the $d_{xz}$ and $d_{yz}$ bands, with enhanced $Q_a$ intensity consistent with dominant $d_{yz}$ orbital occupation in the nematic phase[38]. With increasing electron doping in the underdoped regime, the SDW-folded pockets shift toward the BZ boundary (Fig. 3j), causing $Q_a$ to increase and approach 1/4 ($2\pi/a_{Fe}$). This establishes the momentum-space condition for the formation of the observed $4a_{Fe}$ charge stripe (Fig. 2b,e). Upon further doping into the superconducting regime, both antiferromagnetic and nematic orders are suppressed, eliminating electron-hole hybridization. Instead, hybridization between the two crossed elliptical electron pockets in the folded BZ (two-Fe unit cell) becomes dominant, giving two inner and outer pockets in Fig. 3k. Evidently, the intrapocket scattering within the inner electron pockets offers a straightforward account for the closed, square-shaped QPI features in Supplementary Fig. 11b[59].

Having established that the charge-stripe order originates from FS nesting associated wtih SDW-driven FS reconstruction, we now turn to the strain-induced superconducting puddles in CFCA films. It has been well established that external pressure suppresses both antiferromagnetic and nematic orders in $CaFe_2As_2$ (CFA) by reducing the As anion height within the FeAs plane, thereby stabilizing a nonmagnetic collapsed-tetragonal (CT) phase (Fig. 1b)[60,61]. In the present epitaxial films, the in-plane tensile strain imposed by the $SrTiO_3$ substrate (lattice constant 3.905 Å) on CFCA ($\leq 3.88$ Å) induces an effective out-of-plane compressive strain. This strain promotes the formation of nanoscale collapsed superconducting puddles in the CFCA films (Supplementary Fig. 6), where the $c$-axis lattice constant is reduced by ~ 1.0 Å (Supplementary Fig. 5c), consistent with expectations for the CT phase[61]. To disentangle the effects of epitaxial strain from chemical doping, we investigate CFA films with systematically varied film thicknesses, for which the tensile strain increases as the film becomes thinner. As shown in Fig. 4a-c, nematic domain walls become progressively distorted in 6-UC films and are completely suppressed when the film thickness is reduced to 3 UC and below, signaling the emergence of a CT phase. This transition is accompanied by the onset of superconductivity, evidenced by the appearance of temperature-dependent superconducting gaps (Fig. 4d,e) and a pronounced reduction of the zero-bias conductance (defined as the ratio of conductance at zero energy to that at the coherence energy, i.e., $g(r, E_F)/g(r, E_{coh})$ in Fig. 4f). Remarkably, in the 1-UC CFA film, the superconducting gap closes at ~18 K (Fig. 4e), comparable to that of optimally doped CFCA (Fig. 1f). Unlike the chemically doped case, however, no charge-stripe order is observed across the strain-driven nematic-to-superconducting transition (Fig. 2a and Fig. 4a-c). Instead, both the superconducting puddles (Supplementary Fig. 6a,b) and the ultrathin CFA films (Fig. 4b,c) exhibit $C_4$ symmetric patterns identical to those of the chemically doped superconducting phase.

The unique observation of charge-stripe order in the non-collapsed, underdoped CFCA films, together with its characteristic contrast reversal across the VHS energy, points to a correlation-driven mechanism. In this picture, the collinear antiferromagnetic correlations, electronic nematicity, and the saddle-point VHS collectively promote the formation of a translational-symmetry-breaking charge-ordered state. By directly visualizing this unidirectional electronic state that precedes superconductivity, the present work fills a missing element in the phase diagram of IBSs. The similarity with cuprates suggests that charge ordering represents a universal electronic instability in high-$T_c$ superconductors, emerging from the evolution of the correlated electronic structure with doping rather than from material-specific details alone. The contrasting effects of chemical doping and epitaxial strain – two independent tuning knobs – further demonstrate that the stripe order is not merely a competing instability, but rather a direct spectroscopic manifestation of the correlated normal state proximal to superconductivity. From a methodological perspective, the ability to resolve and manipulate charge-stripe textures in CFCA at the atomic scale provides a powerful platform for studying how emergent charge order shapes low-energy excitations and for testing microscopic descriptions of intertwined quantum phases in correlated materials.

These findings carry several critical implications for charge ordering in correlated superconductors. First, although charge order appears across multiple correlated materials, its microscopic form is not universal. The stripe order observed here differs fundamentally from the checkerboard-like charge order commonly reported in cuprates, likely reflecting the unique roles of collinear antiferromagnetic and nematic correlations in iron-based systems. This contrast highlights the need to further clarify how these orders affect the onset, orientation, and stability of translational-symmetry-breaking charge order. Second, in CFCA, charge stripes emerge prior to superconductivity, indicating a competitive interplay between the two orders. This behavior contrasts with recent observations in a sister material Ba(Fe$_{1-x}$Co$_x$)$_2$As$_2$, where charge stripes are hidden in superconducting samples but become observable only after superconductivity is locally suppressed by magnetic fields[62]. These distinctions indicate a complex and system-dependent nature of charge order-superconductivity interplay even within closely related materials. Finally, the role of epitaxial strain found here differs qualitatively from that in LiFeAs and FeSe-derived systems, where strain can induce charge order[45,46]. In contrast, no charge order is stabilized across the entire strain range explored in the CFCA films (Fig. 4), pointing to distinct microscopic strain-coupling mechanisms that remain to be understood. Nevertheless, the unique combination of epitaxial strain engineering and precise doping control demonstrated here provides a decisive experimental framework for disentangling the underlying mechanisms governing competing electronic orders in high-$T_c$ superconductors.

**Methods**

**Sample growth.** High-quality and atomically flat Ca(Fe$_{1-x}$Co$_x$)$_2$As$_2$ (CFCA) films with precisely controlled Co doping levels were epitaxially prepared on 0.5wt% Nb-doped SrTiO$_3$(001) substrates under ultra-high vacuum (UHV) conditions using molecular beam epitaxy (MBE), maintaining a base pressure better than $2.0 \times 10^{-10}$ Torr. To achieve atomically flat and clean surface for epitaxy, the SrTiO$_3$ substrates were first degassed at 600°C for 3 hours, followed by annealing at 1250°C for 20 min. Fluxes of high-purity Ca (99.99%), Fe (99.9999%) and Co (99.9999%) were exactly calibrated by a standard quartz crystal microbalance (Inficon SQM160H) before growth. The films were then grown at a substrate temperature of 500°C by co-evaporating metal sources (Ca, Fe and Co) from Knudsen cells at controlled stoichiometry under an As-rich condition (> $10^{-7}$ Torr), with a typical growth rate of ~ 0.17 unit cells per minute. The Co doping level and film thickness were determined based on the Co/Fe flux ratio and the growth duration.

**STM measurement.** STM measurements were performed using a commercial Unisoku USM 1500 system, capable of applying a perpendicular magnetic field up to 8 T. The base pressure was maintained below $2.0 \times 10^{-10}$ Torr. Polycrystalline PtIr tips, conditioned by electron-beam bombardment in UHV and calibrated on MBE-grown Ag/Si(111) films, were used throughout the experiments. All STM topographies were recorded in constant current mode, while the d$I$/d$V$ spectra and differential conductance maps $g(\bm{r}, V)$ were acquired using a standard lock-in technique with a small *a.c.* modulation voltage (typically ~ 1/50 of the applied sample bias) at a frequency of $f$ = 913 Hz. STM measurements were performed at 4.4 K unless otherwise stated.

**Transport measurements**: After *in-situ* STM measurements, the samples were taken out of the ultra-high vacuum chamber for electrical measurement. The electrical resistivity was measured via a four-terminal configuration ($I$ = 1 $\mu$A) in a commercial physical property measurement system (PPMS).

**Two-dimensional lock-in analysis.** To extract the amplitude of charge stripes from STM spectroscopic maps, we utilized a standard two-dimensional lock-in technique[11]. For example, an STM conductance map $g(\bm{r}, E)$ can be regarded as a superposition of periodic charge stripe modulations, quasiparticle interference modulations, and noise signals. By applying a Fourier transform to $g(r, E)$, one can effectively distinguish potential modulation signals. This methodology is expressed as

$$g(r, E) = \sum_Q A_Q(r, E) e^{iQ \cdot r} \tag{1}$$

$$g(q, E) = \sum_Q A_Q(q - Q, E), \tag{2}$$

where $A_Q(q - Q, E)$ represents the amplitude distribution function of a modulation at wave vector $Q$ for energy $E$. The amplitude $A_Q(q - Q, E)$ of an ideal plane wave would concentrate at a single pixel while various factors such as thermal fluctuations, tip drift, lattice strain and defects would cause inhomogeneous distribution, broadening the single peak. To extract the modulation signal at wave vector $Q$, we utilize a standard lock-in method which involves multiplying $g(q, E)$ (Fourier transform of $g(r, E)$) by a Gaussian packet centered at $Q$, followed by an inverse Fourier transform.

$$g_Q(r, E) = F^{-1}\left[g(q, E) \frac{1}{\sqrt{2\pi}\sigma_Q} \exp\left(-\frac{(q - Q)^2}{2\sigma_Q^2}\right)\right]. \tag{3}$$

Here the cutoff length $\sigma_Q$ of the Gaussian packet in the $q$ space should satisfy

$$\lambda_Q < \sigma_Q \ll \delta_Q, \tag{4}$$

where $\lambda_Q$ represents the characteristic broadening length of $A_Q(q - Q, E)$ and $\delta_Q$ is the distance between $Q$ and the nearest wave vector around it. The extracted single wave vector map $g_Q(r, E)$ can be described as

$$g_Q(r, E) = A(r, Q, E) \exp[-i(Q \cdot r + \phi(r, Q, E))]. \tag{5}$$

Thus we can derive the amplitude map $A(r, Q, E)$ and phase map $\phi(r, Q, E)$ from the inverse Fourier transform

$$A(r, Q, E) = |g_Q(r, E)| \tag{6}$$

$$\phi(r, Q, E) = -Q \cdot r + i \ln \frac{g_Q(r, E)}{A(r, Q, E)} \tag{7}$$

In addition, one can also obtain the real part of the inverse Fourier transform

$$g(r, Q, E) = A(r, Q, E) \cos(\phi(r, Q, E)), \tag{8}$$

which includes both the magnitude and phase information of the modulations at wave vector $Q$.

**Data availability**

All data that support the findings of this study are available from the corresponding authors upon reasonable request.

**Acknowledgments**

The work is financially supported by grants from the National Key Research and Development Program of China (Grant No. 2022YFA1403100) and the Natural Science Foundation of China (Grant No. 12474130, 12141403 and Grant No. 52388201).

**Author contributions**

C.L.S., X.C.M. and Q.K.X. conceived and supervised the research. Q.J.C., C.C.L. and Y.W.W. grew the samples and performed the STM measurements. Z.X.D. carried out the transport measurements. Q.J.C. and C.C.L. analyzed the experimental data and plotted the figures. Q.J.C. and C.L.S. wrote the manuscript with comments from all of the authors.

**Competing financial interests**

The authors declare no competing financial interests.

**Additional information**

Supplementary information includes 11 figures and the corresponding figure captions.

†Correspondence and request for materials should be addressed to C.L.S. (email: clsong07@mail.tsinghua.edu.cn), X.C.M. (email: xucunma@mail.tsinghua.edu.cn) or Q.K.X. (email: xucunma@mail.tsinghua.edu.cn).

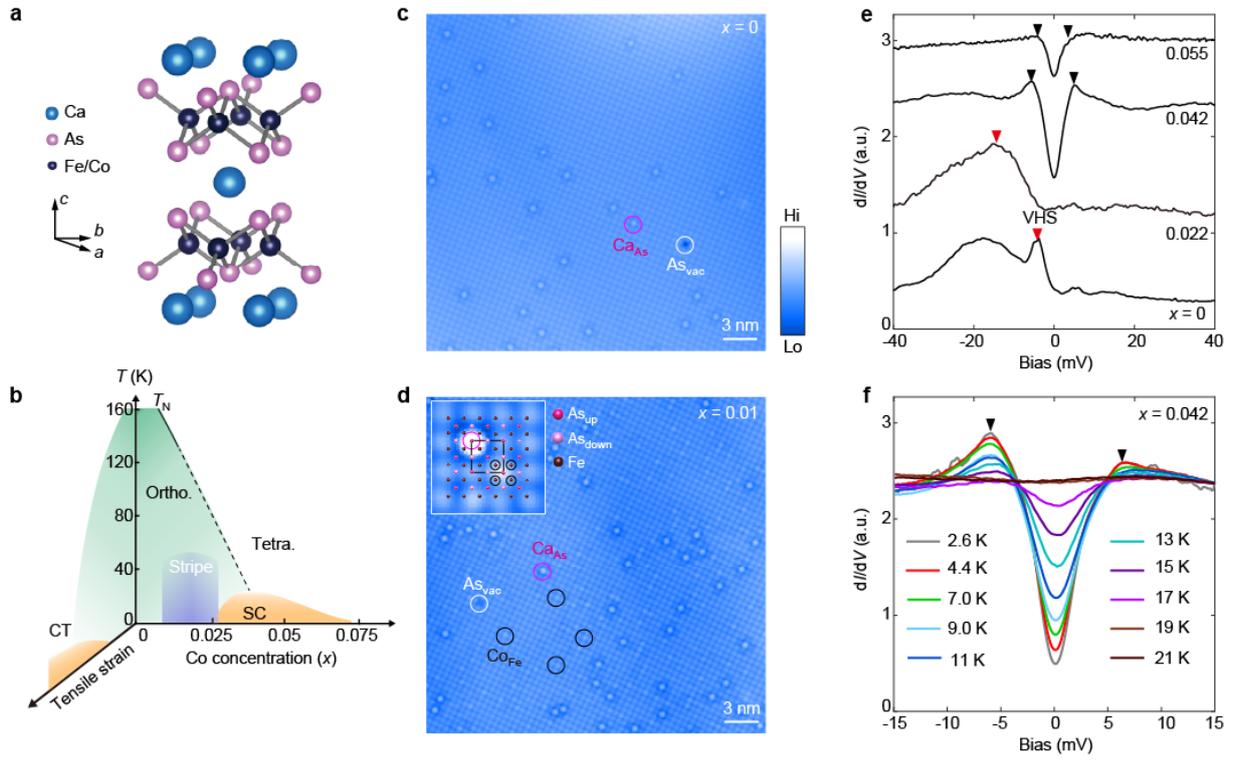

**Fig. 1| Characterization of FeAs planes in Ca(Fe$_{1-x}$Co$_x$)$_2$As$_2$ (CFCA) films. a**, Schematic crystal structure of CFCA. **b**, Electronic phase diagram of CFCA films as a function of Co concentration $x$ and tensile strain. Tetra, tetragonal phase; Ortho, orthorhombic phase; SC, superconducting phase; CT, collapsed tetragonal phase. **c,d**, Atomic-resolution STM images (30 nm × 30 nm, $V$ = 70 mV, $I$ = 0.1 nA) of the parent (**c**) and slightly doped (**d**, $x$ = 0.015) CFCA films. Three types of defects are identified as Ca substitution for As sites (Ca$_{sub}$, circled in magenta), As vacancies (As$_{vac}$, circled in white), and Co substitution at Fe sites (Co$_{Fe}$, circled in black). Inset in **d** shows the atomic model overlaid on the $\sqrt{2} \times \sqrt{2}$ reconstructed lattice (outlined by black box), as well as the Ca$_{sub}$ and Co$_{Fe}$ defects. **e**, Co-doping ($x$) dependence of spatially averaged d$I$/d$V$ spectra at 4.2 K, vertically offset for clarity. Setpoint: $V$ = 40 mV, $I$ = 0.5 nA. The VHS positions and superconducting coherence peaks are marked by red and black triangles, respectively. **f**, Temperature dependence of the superconducting gap in the OPT film. Setpoint: $V$ = 15 mV, $I$ = 0.5 nA.

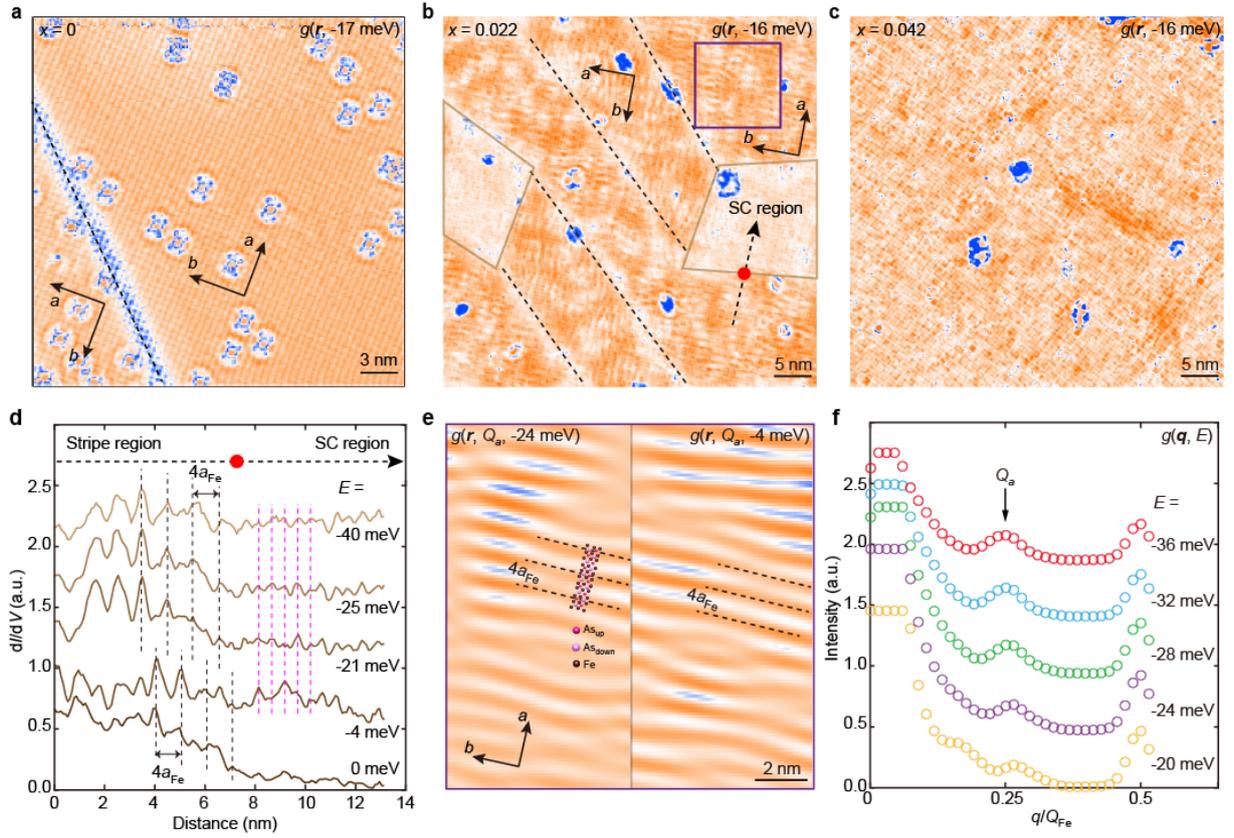

**Fig. 2| Phase evolution of FeAs plane with electron doping. a**, Conductance $g(r, E) \equiv dI/dV(r, E = -17$ meV) map (30 nm × 30 nm, setpoint: $V = 70$ mV, $I = 0.5$ nA) measured in the parent CFA sample (10 UC), revealing nematic domain walls (indicated by a black dashed line). The crystallographic $a$ and $b$ directions are indicated throughout. **b**, $g(r, -16$ meV) map (50 nm × 50 nm, setpoint: $V = 70$ mV, $I = 0.5$ nA) in an UD sample, showing the coexistence of charge-stripe regions and superconducting puddles (outlined by brown boxes). **c**, $g(r, -16$ meV) map (50 nm × 50 nm, setpoint: $V = 70$ mV, $I = 0.5$ nA) in an OPT sample. **d**, Energy-dependent conductance line profiles extracted along the dashed arrow in **b**. Vertical black and magenta dashed lines indicate apparent modulations induced by the charge order in the stripe region (left side) and the $\sqrt{2} \times \sqrt{2}$ reconstructed lattice in the superconducting region (right side), while the boundary is marked by a red spot. **e**, Inverse Fourier transform of $g(q, E)$ at the charge-stripe wave vector $Q_a$, obtained from the stripe region (outlined by the black square in **b**) above (right) and below (left) the VHS energy. **f**, Representative line profiles along the Γ-Y direction in $g(q, E)$.

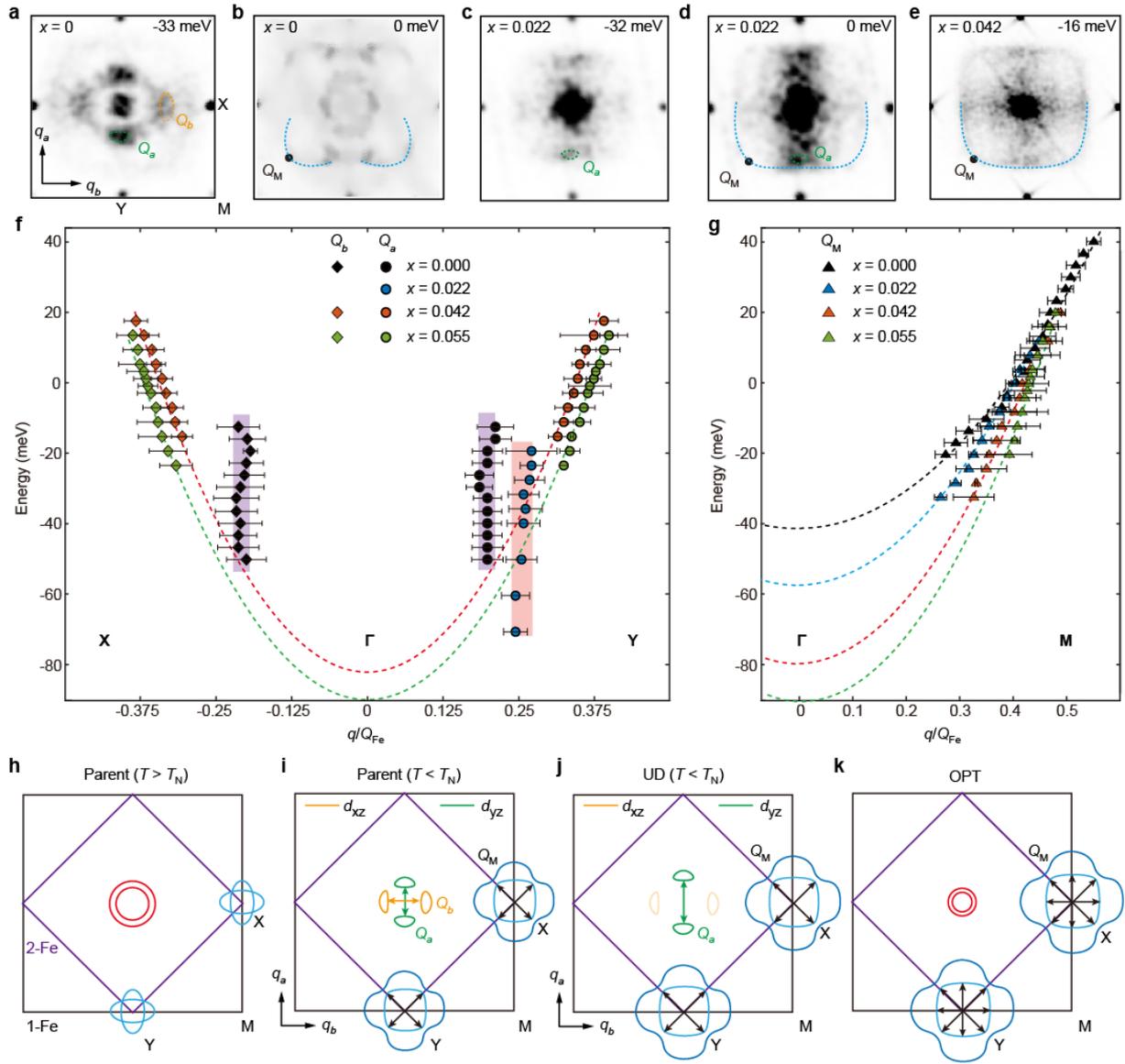

**Fig. 3| Band structure and origin of the stripe order. a-e**, Representative QPI patterns of CFCA samples with varying chemical doping. Anisotropic scattering vectors along the Fe-Fe direction ($Q_a/Q_b$) are marked by green and orange ellipses, respectively, while isotropic scattering along the As-As direction ($Q_M$) is partly highlighted by the blue arcs. **f,g**, Energy and doping dependence of the scattering vectors $Q_a$ and $Q_b$, as well as dispersive QPI patterns along the high-symmetry (**f**) ΓX(Y) and (**g**) ΓM directions in CFCA samples. Dashed lines show parabolic fits of the *E-q* dispersions in the OPT and OD samples. **h**, Schematic FS in the high-temperature tetragonal phase of CFCA films. The Brillouin zones of 1-Fe and 2-Fe are outlined by the black and purple squares, respectively. **i,j** Schematic FS in the antiferromagnetic and nematic phases. The anisotropic scattering vectors at $Q_a(Q_b)$ are marked by green (orange) arrows, while the nearly isotropic scattering vector is marked by the black arrows. **k**, Schematic FS in the OPT CFCA sample.

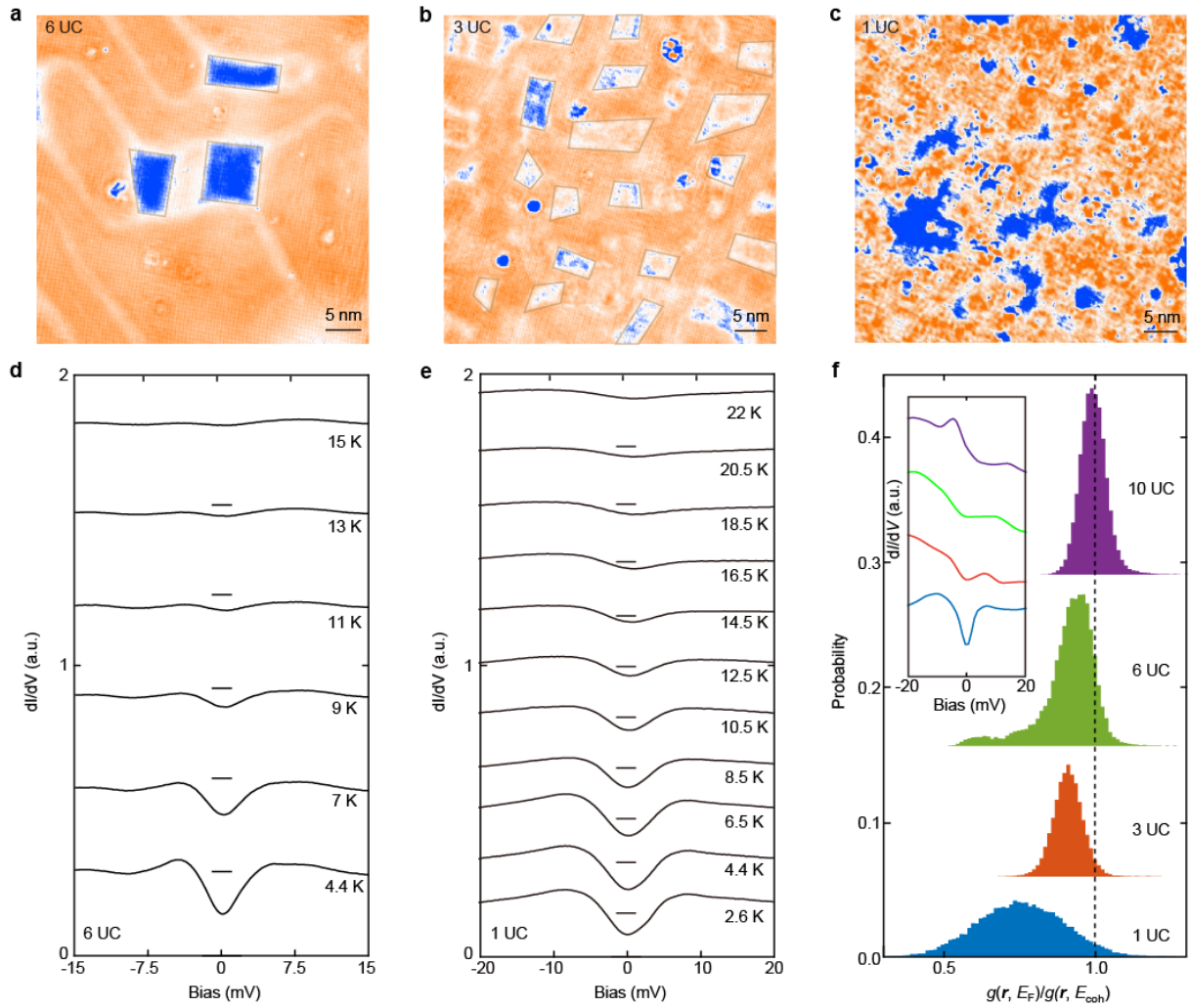

**Fig. 4| Superconductivity and absence of charge-stripe order in collapsed CFCA films. a-c**, Zero-bias conductance maps of parent CFA films with thicknesses of 6 UC (**a**), 3 UC (**b**) and 1 UC (**c**). Regions exhibiting the CT phase are outlined by brown boxes. Setpoint: $V = 70$ mV, $I = 0.5$ nA. **d,e**, Temperature dependence of the superconducting gap measured within the collapsed regions (brown boxes) of 6-UC (**d**) and 1-UC (**e**) CFA films. Setpoint: $V = 70$ mV, $I = 0.5$ nA. **f**, Histograms of the relative zero-bias conductance extracted from the full fields of view in **a-c**. Inset shows spatially averaged $dI/dV$ spectra collected from the entire fields of view in **a-c** and Fig. 2a (10 UC film).

*Supplementary Materials for*

**Intrinsic translational symmetry-breaking charge stripes in underdoped iron pnictides**

Qiang-Jun Cheng[1], Cong-Cong Lou[1], Yong-Wei Wang[1], Ze-Xian Deng[1], Xu-Cun Ma[1,2,†], Qi-Kun Xue[1,2,3,4,5,†], Can-Li Song[1,2,†]

[1]*Department of Physics and State Key Laboratory of Low-Dimensional Quantum Physics, Tsinghua University, Beijing 100084, China*

[2]*Frontier Science Center for Quantum Information, Beijing 100084, China*

[3]*Shenzhen Institute for Quantum Science and Engineering and Department of Physics, Southern University of Science and Technology, Shenzhen 518055, China*

[4]*Beijing Academy of Quantum Information Sciences, Beijing 100193, China*

[5]*Hefei National Laboratory, Hefei 230088, China*

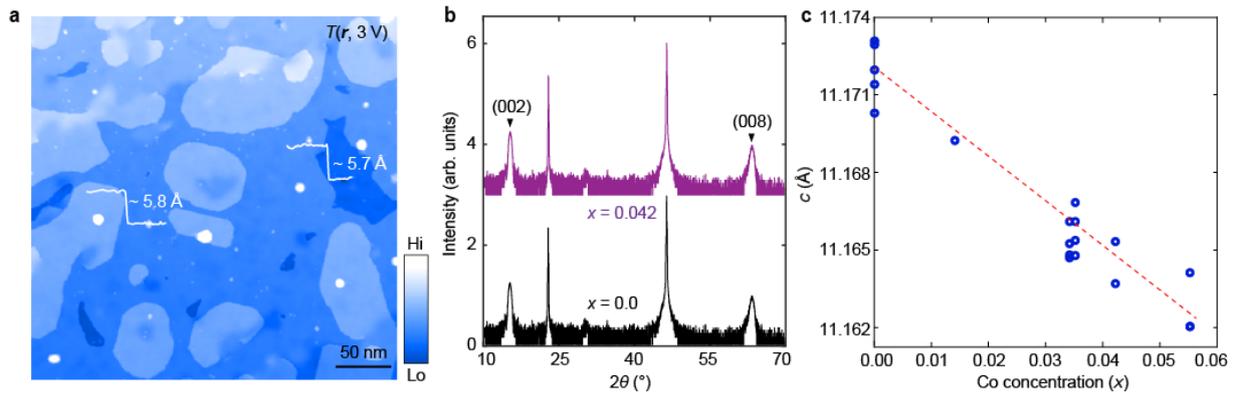

**Supplementary Fig. 1. Epitaxial CFCA films with varying doping levels. a**, Large-area STM topography (300 nm × 300 nm, $V$ = 3.0 V) acquired on an undoped $CaFe_2As_2$ film with a thickness of 10 UC. Two representative line profiles extracted across half-unit-cell steps are shown. **b**, XRD patterns of 10-UC parent and optimally doped CFCA films. The (00n) diffraction peaks of CFCA are labelled. **c**, Out-of-plane lattice constant $c$ as a function of Co concentration $x$. The dashed line denotes a linear fit to the data.

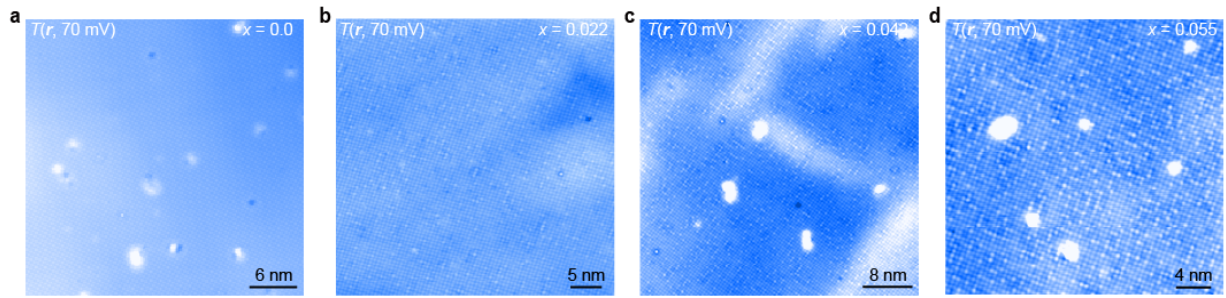

**Supplementary Fig. 2. Co dopants in the CFCA films. a-d**, Atomic-resolution STM topographies $T(r, 70$ mV) of CFCA films in the parent (**a**), underdoped (**b**), optimally doped (**c**) and overdoped (**d**) regimes. The density of bright protrusions, attributed to Co substitutions at Fe sites, increases systematically with the Co doping concentration $x$.

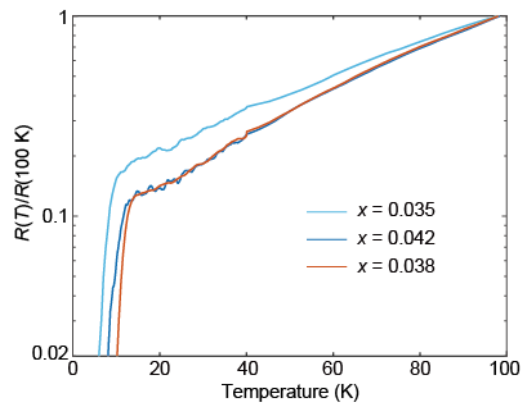

**Supplementary Fig. 3. Temperature-dependent resistivity of 10-UC CFCA films**. All curves have been normalized to the resistivity at 100 K and plotted on a semi-logarithmic scale.

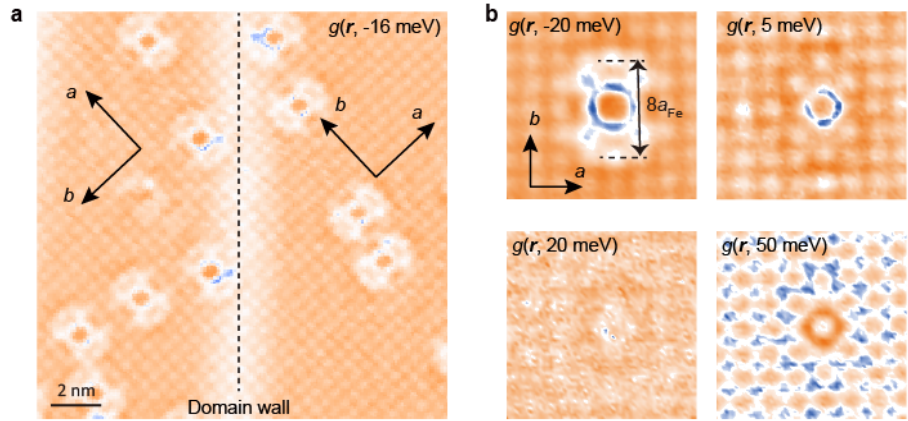

**Supplementary Fig. 4. Dumbbell-shaped defects. a**, Conductance map $g(r, -16\ \text{meV})$ revealing dumbbell-shaped defect states. The dashed line marks a nematic domain boundary, across which the orthorhombic $a$- and $b$-axes interchange. **b**, Zoomed-in views of the dumbbell-shaped defects at selected energies, exhibiting a characteristic spatial extent of $\sim 8a_{Fe}$ and a pronounced energy-dependent evolution of the defect states.

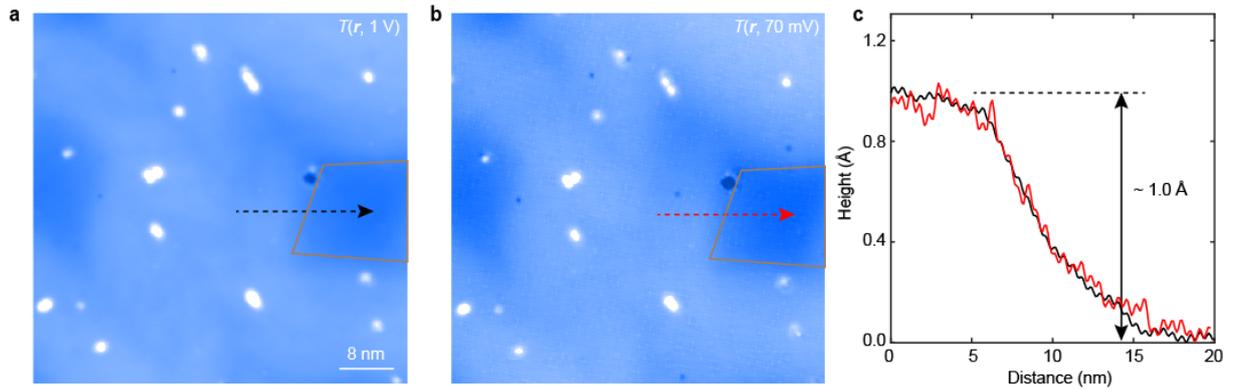

**Supplementary Fig. 5. Collapsed regions in UD samples. a,b**, STM topographies of an UD sample taken in the same field of view as Fig. 2**b**, revealing the presence of collapsed regions (brown boxes). The images are acquired at applied sample biases of 1 V (**a**) and 70 mV (**b**). **c**, Height line profiles across the collapsed regions, taken along the black and red arrows in **a** and **b**, respectively.

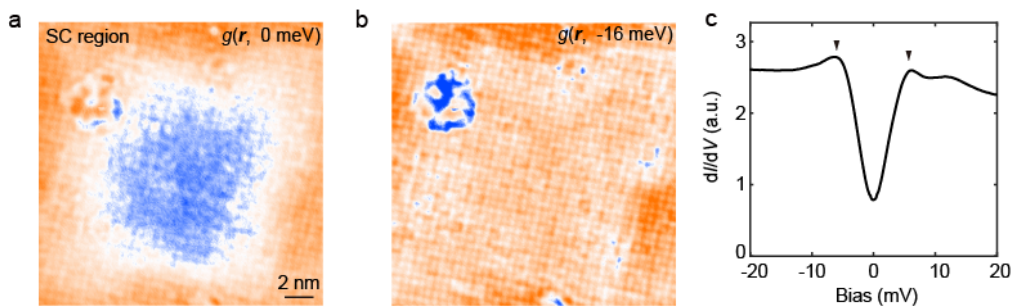

**Supplementary Fig. 6. Superconducting puddles in UD CFCA films. a,b**, Conductance maps $g(r, E)$ (20 nm × 20 nm, setpoint: $V = 70$ mV, $I = 0.5$ nA) of a superconducting puddle, measured at 0 meV (**a**) and -16 meV (**b**), respectively. **c**, Spatially averaged d$I$/d$V$ spectrum measured within the superconducting puddle, with the superconducting coherence peaks marked by black triangles. Setpoint: $V = 20$ mV, $I = 0.5$ nA.

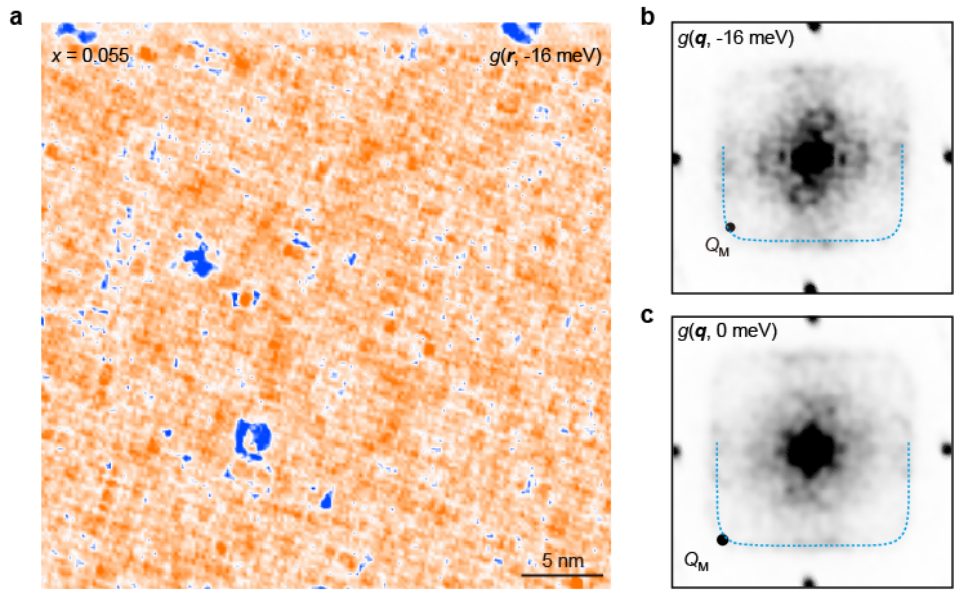

**Supplementary Fig. 7. Electronic structure of OD CFCA films. a**, Representative conductance map $g(r, -16\ \text{meV})$ acquired on an OD CFCA film. Setpoint: $V = 70$ mV, $I = 0.5$ nA. **b,c**, Fast Fourier transforms of the conductance maps $g(r, -16\ \text{meV})$ (**b**) and $g(r, 0\ \text{meV})$ (**c**) of the OD sample. The isotropic scattering is partially highlighted by the blue dashed arcs.

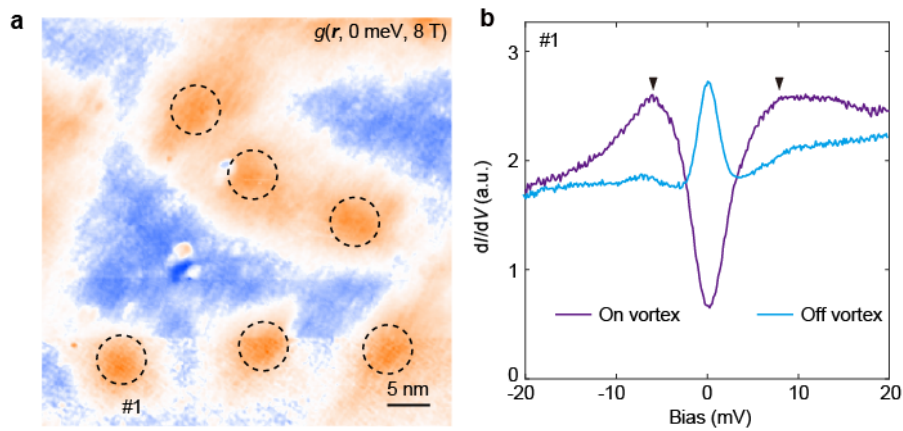

**Supplementary Fig. 8. Superconductivity of OPT CFCA films. a**, Zero-bias conductance map $g(r, 0\ \text{meV})$ of an OPT CFCA film ($x \sim 0.042$) measured under an applied magnetic field of 8 T. Magnetic vortices are outlined by dashed circles. **b**, Representative $dI/dV$ spectra acquired at vortex core #1 and away from the vortex (off vortex). Setpoint: $V = 20$ mV, $I = 0.5$ nA.

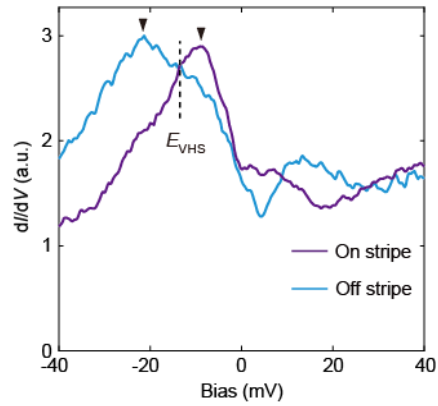

**Supplementary Fig. 9. Spectral modulation in the charge-stripe region of UD samples.** Representative d$I$/d$V$ spectra acquired on and off the charge-stripe region in Fig. 2**b**, showing a modulated asymmetric peak near the VHS energy. Setpoint: $V$ = 40 mV, $I$ = 0.5 nA.

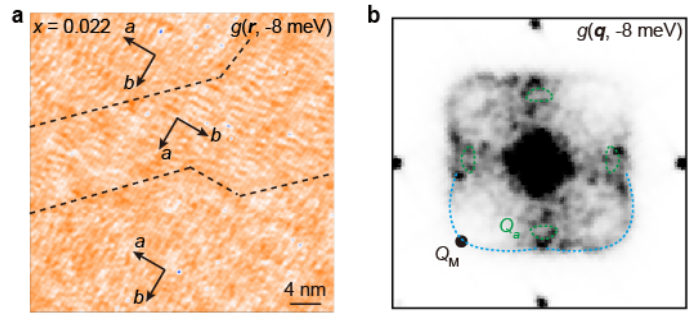

**Supplementary Fig. 10. QPI measurements in UD CFCA films. a**, Large-scale conductance map $g(r, -8$ meV) taken in a different region of the UD CFCA film ($x$ = 0.022), revealing orthogonal nematic domains marked by black dashed lines. Setpoint: $V$ = 70 mV, $I$ = 0.5 nA. **b**, Fast Fourier transform of the conductance map $g(r, -8$ meV) in **a**, calculated over the entire field of view.

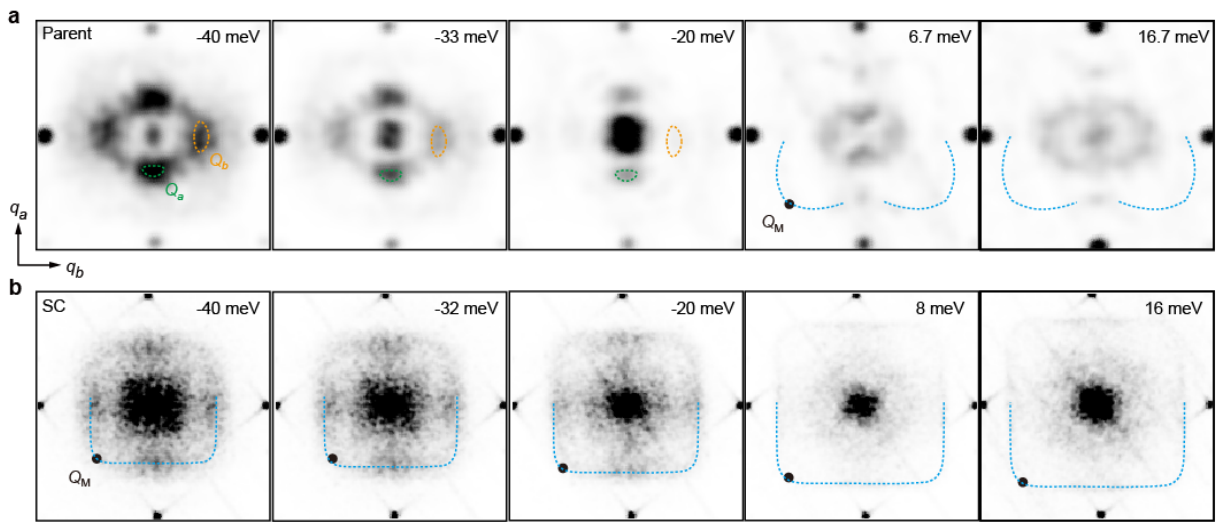

**Supplementary Fig. 11. Comparison of the electronic structure between parent and SC samples. a**, A series of QPI images obtained from a single nematic domain of the parent CFA sample at various energies. **b**, QPI images acquired from superconducting CFCA samples at energies comparable to those shown in **a**.